\documentclass{webofc}
\usepackage{natbib}
\usepackage[varg]{txfonts}   
\begin{document}
\title{Systematic benchmarking of HTTPS third party copy on 100Gbps links using XRootD}
%
%

\author{
        \firstname{Edgar} \lastname{Fajardo}\inst{1}\fnsep\thanks{\email{emfajard@ucsd.edu}}
        \and
        \firstname{Aashay} \lastname{Arora}\inst{1}\fnsep\thanks{\email{aaarora@ucsd.edu}}
        \and
        \firstname{Diego} \lastname{Davila}\inst{1}\fnsep\thanks{\email{didavila@ucsd.edu}}
        \and
        \firstname{Richard} \lastname{Gao}\inst{1}\fnsep\thanks{\email{rgao@ucsd.edu}}
        \and
        \firstname{Frank} \lastname{W\"urthwein}\inst{1}\fnsep\thanks{\email{fkw@ucsd.edu}}
        \and
        \firstname{Brian} \lastname{Bockelman}\inst{2}\fnsep\thanks{\email{bbockelman@morgridge.org}}
}

\institute{University of California San Diego, 9500 Gilman Dr, La Jolla, CA 92093
\and Morgridge Institute, 330 N Orchard St, Madison, WI 53715 
          }

\abstract{%
  The High Luminosity Large Hadron Collider provides a data challenge. The amount of data recorded from the experiments and transported to hundreds of sites will see a thirty fold increase in annual data volume. A systematic approach to contrast the performance of different Third Party Copy (TPC) transfer protocols arises. Two contenders, XRootD-HTTPS and the GridFTP are evaluated in their performance for transferring files from one server to another over 100Gbps interfaces. The benchmarking is done by scheduling pods on the Pacific Research Platform Kubernetes cluster to ensure reproducible and repeatable results. This opens a future pathway for network testing of any TPC transfer protocol.
}
\maketitle
\section{Introduction}
\label{intro}
\begin{figure}[h]
\centering
\includegraphics[width=10cm,clip]{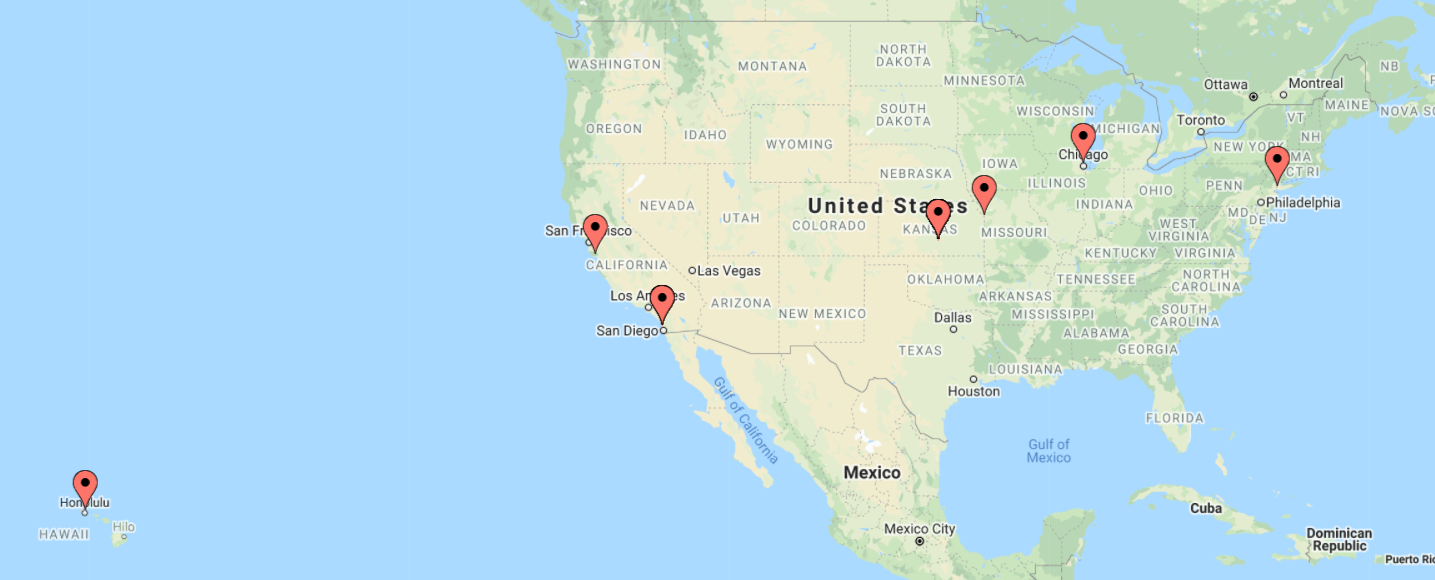}
\caption{Location of 100Gbps capable servers in the PRP Nautilus cluster.}
\label{fig-prp-location}       
\end{figure}

The Worldwide LHC Computing Grid (WLCG)\cite{wlcg} is evaluating different file transfer protocols for the High Luminosity Large Hadron Collider (HL-LHC) era. In the HL-LHC data volumes are expected to see a twenty to forty fold volume increase. At the same time the grid computing centers have augmented their network bandwidth with several of them connected at 100Gbps. This presents a three-fold challenge on software and infrastructure: how to deploy infrastructure services that will cope with  the increased data volumes, how to efficiently use the Wide Area Network (WAN) links between sites for file transfers and how to systematically measure the performance of different protocols on these links.
The \textit{Pacific Research Platform} (PRP) runs a worldwide distributed Kubernetes cluster in which researchers can deploy any application on any of its nodes in the form of Docker containers. Of interest to this work are the more than twenty 100Gbps capable servers located around North America (see Fig \ref{fig-prp-location}). An opportunity arises to deploy any file transfer protocol on these nodes and measure its actual server to server performance and the sensitivity to latency, given the highly distributed nature of the available nodes.

Two of the most used file transfer protocols in the WLCG are GridFTP\cite{gridftp} and XRootD-HTTPS \cite{xrootdTPC,dorigo2005xrootd}. In this project both are evaluated for their throughput and latency sensitivity. Different Docker containers are created out of their corresponding RPM packages maintained by the Open Science Grid (OSG)\cite{osg}. 

\begin{figure}[h]
\centering
\includegraphics[width=8cm,clip]{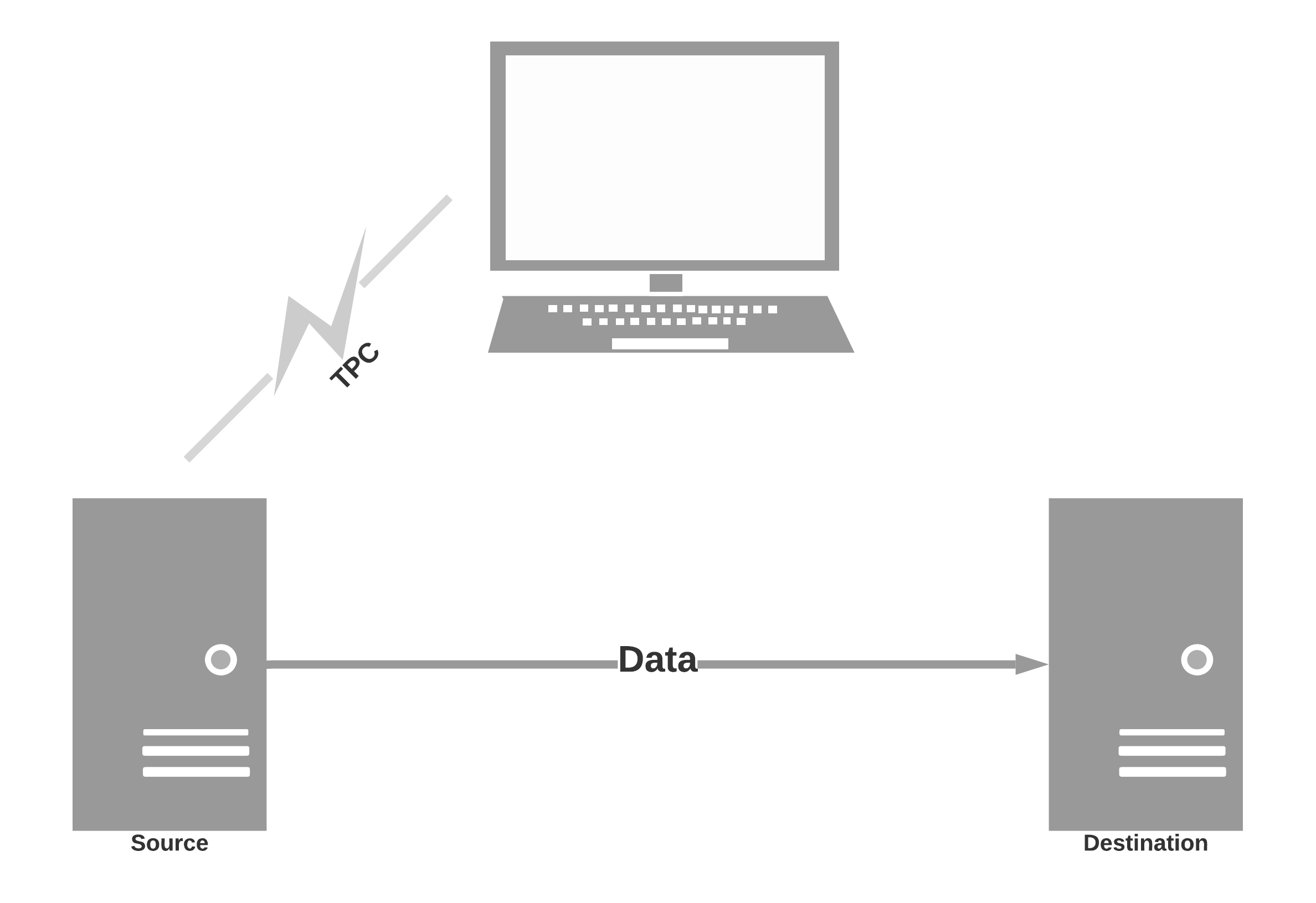}
\caption{Diagram of Third Party Copy transfer}
\label{fig-tpc}       
\end{figure}

Both of these protocols support Third Party Copy (TPC), this mechanism allows a "lightweight" client to start a transfer between a source and destination as opposed to the client streaming the file from the source and then uploading it to the destination (see Figure \ref{fig-tpc}).

To fully understand the performance of these protocols the first step is to understand the limitation of the Docker containers without involving the network (see Section \ref{sec-preliminary}). Then the protocol limitations  are tested on ideal conditions: low latency, memory based file system (see Section \ref{sec-low-latency-limit}). Finally a systematic approach is taken to run file transfers at all locations and with different parameters protocol, number of streams, etc (see Sections \ref{sec-maddash} and \ref{sec-full-benchmarking}).

\section{Preliminary Benchmarking}
\label{sec-preliminary}
\begin{figure}[h]
\centering
\frame{\includegraphics[width=8cm,clip]{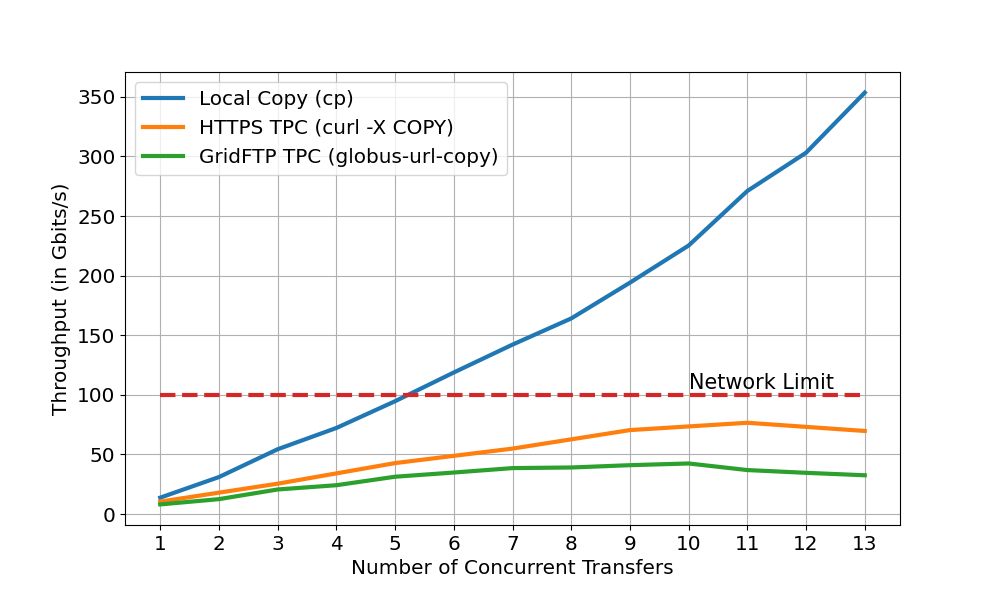}}
\caption{Single pod reading/writing speeds for copy (cp), XRootD-HTTPS TPC (curl) and GridFTP TPC (globus-url-copy),}
\label{mmc} 
\end{figure}

\begin{figure}[h]
\centering
\frame{\includegraphics[width=8cm,clip]{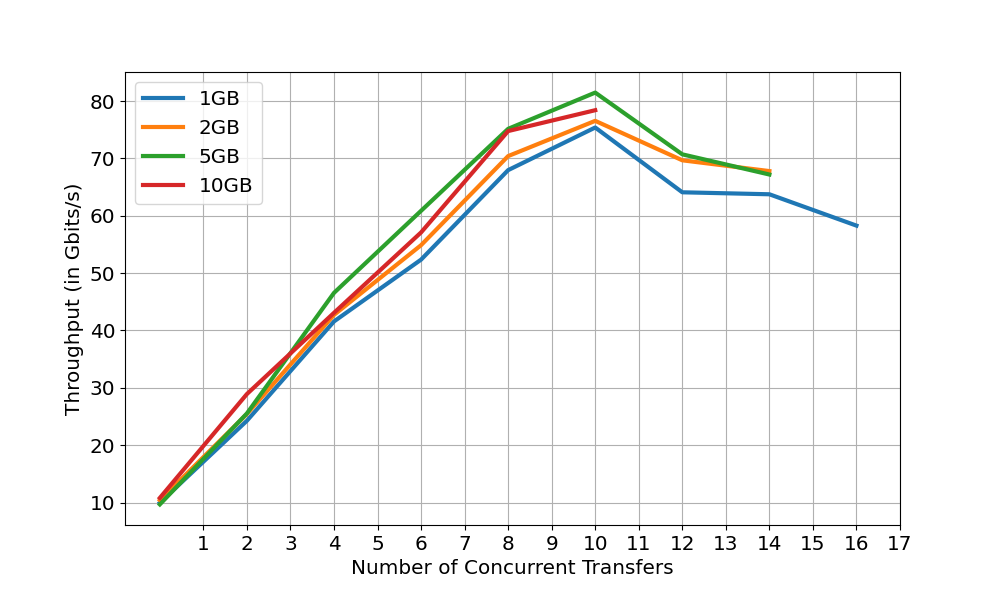}}
\caption{XRootD-HTTPS TPC Throughput for concurrent transfers done using 1GB (blue), 2GB (yellow), 5GB (green) and 10GB (red) files.}
\label{fig-filesize-comp}
\end{figure}

The read/write speeds from memory and disk are tested before the full WAN file transfers. This first step ensures that the limitations found when transferring files are inherent to the protocols and their implementations rather than the hardware where they are deployed. Concurrent file copies are run between two memory mounts in the same pod and show that six concurrent copies are enough to reach write speeds close to one hundred Gigabits per second (Gbps) (see blue line on Figure \ref{mmc}).
Similar tests are conducted to assess the overhead introduced by the protocol implementation services running on pods and reading and writing to POSIX memory 
file mounts. Multiple concurrent transfers are run between the two mounts, each one running an instance of the service. These local transfers use the pod's loopback network interface in order to rule out WAN limits. This showcases that the maximum throughput achievable is 82 Gbits/sec at 11 concurrent transfers using XRootD-HTTPS and about 42 Gbits/sec at 9 concurrent transfers using GridFTP (see orange and green lines on Figure \ref{mmc}).
The aforementioned tests are run using files of various sizes (see Figure \ref{fig-filesize-comp}) to propose an optimal file size for the actual measurement. It is observed that the variance in the throughput due to changing the file size is minimal. Therefore, in order to minimize cluster resource usage, 1GB files are designated as the metric.
\subsection{Low Latency Limits}
\label{sec-low-latency-limit}
\begin{figure}[h]
\centering
\frame{\includegraphics[width=13cm,clip]{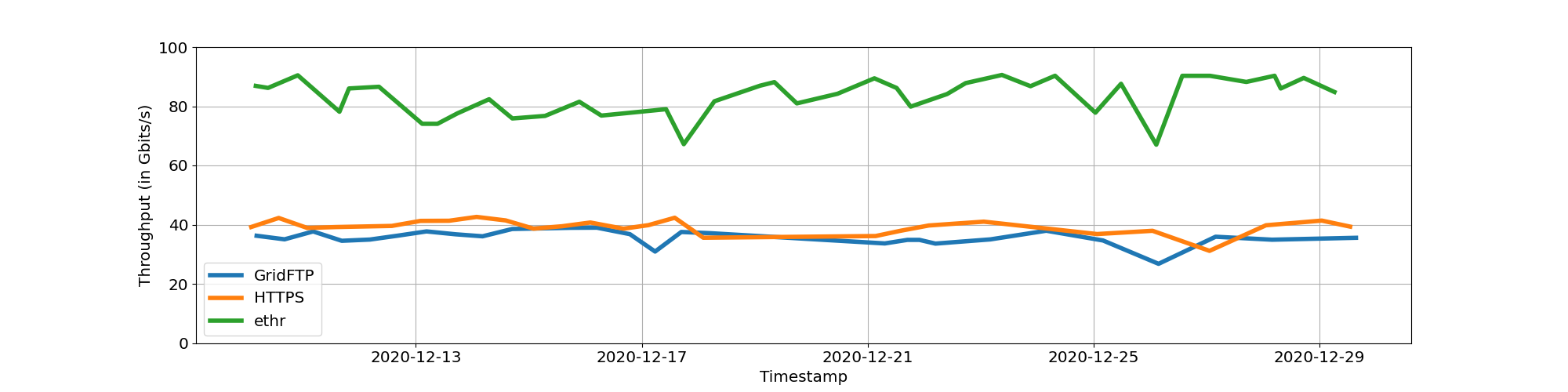}}
\caption{Throughput between two nodes with low (<1ms) latency measured XRootD-HTTPS TPC (orange), GridFTP TPC (blue) and ethr(green) using eight streams.}
\label{fig-lowlatency-throughput}
\end{figure}

After the single pod testing on Section \ref{sec-preliminary} the natural next step is to involve more than one node and measure the results. This is done in a controlled scenario: two nodes in the same computer room. This step produces the baseline results for TPC transfers over the WAN. The results can be seen in figure \ref{fig-lowlatency-throughput}. All depicted transfers use eight streams to ensure uniformity.

This shows the results of two weeks of twice a day transfers between the pods. At low latency using eight streams XRootD-HTTPS barely outperforms GridFTP and both are bellow forty percent less than the total link capacity measured by ethr\cite{ethr}. This is consistent with the results previously obtained in \cite{chep2019}.
\section{Kubernetes and Maddash deployment}
\label{sec-maddash}
Once the baseline limitations of the protocols are understood in the one pod and two pod scenario the next step is to do systematic measurements in a wide range of hosts. The PRP Nautlius Kubernetes cluster is used to deploy one pod per protocol on each 100Gbps capable node. Kubernetes DaemonSet service allows for the deployment of a large number of pods on all selected nodes simultaneously \cite{tpc-bench}, thereby eliminating the need for a separate definition for each host. The pods use host network (as opposed to virtual network) to ensure the endpoint IP address does not change and to avoid potential slowdowns. To simulate the real-world use case of TPCs over HTTPS, the OSG Certificate Authority (CA) generator \cite{osg-ca-gen} is used to simulate a certificate authority and create an X509 certificate for every pod and node. 

\begin{figure}[h]
\centering
\frame{\includegraphics[width=12cm,clip]{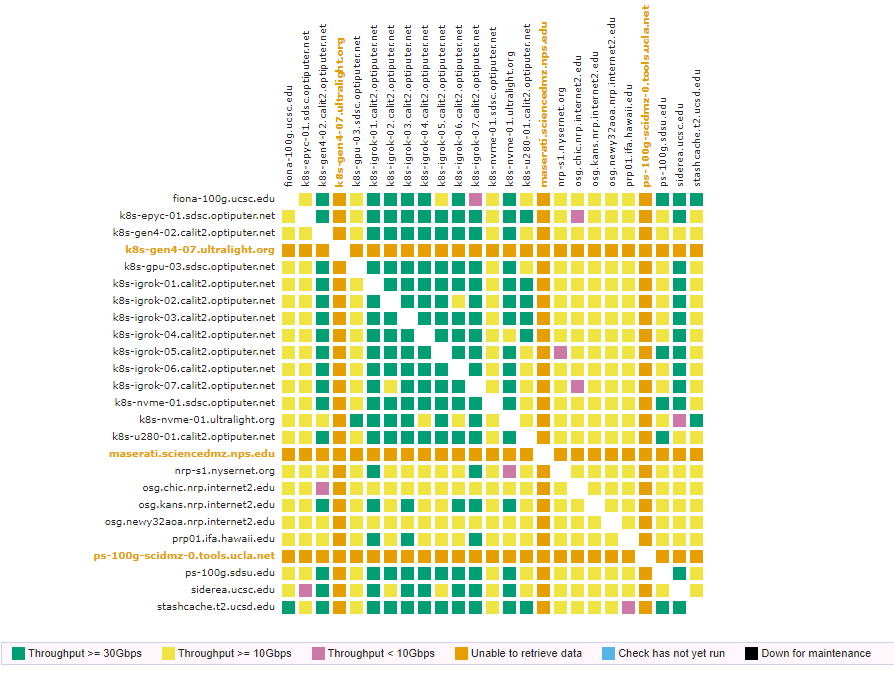}}
\caption{Maddash of the combined throughput results in all 100Gbps hosts in the Pacific Research platform. }
\label{fig-maddash}
\end{figure}

A "master" pod responsible for orchestrating the transfers between the servers is deployed. The throughput is calculated then by dividing the total data transferred by the average time taken for one transfer. The transfers are scheduled every twelve hours and the measurements are recorded over a three month period. In order to understand the relation between the throughput and number of streams the transfers are performed twice every twelve hours. First one uses eight streams and the second one uses a single stream where each "measurement" consists of eleven one gigabyte file concurrent transfers for each source-destination pair.
After each set of measurements, the results are aggregated by the master pod and sent to a Django database where they can be pulled and visualized on a MadDash\cite{maddash} dashboard as a two-dimensional grid (see Figure \ref{fig-maddash}). MadDash is a tool commonly used by network engineers to monitor network performance, hence it was a natural selection to show the results of network file transfer protocols.

\section{Final Results}
\label{sec-full-benchmarking}
\begin{figure}[h]
\centering
\frame{\includegraphics[width=8cm,clip]{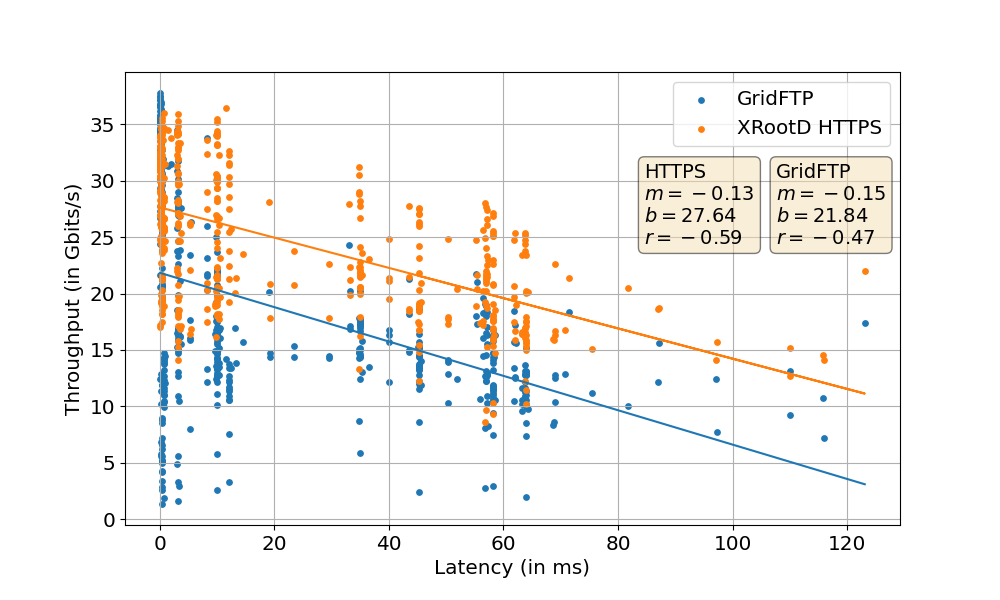}}
\caption{Comparison of throughput using HTTPS and GridFTP (eight streams) as a function of latency}
\label{fig-8-stream-latency}
\end{figure}

Once several weeks of data are stored in the database it is possible to answer what is the average behaviour of the protocols in different scenarios. Figure \ref{fig-8-stream-latency} shows that once all data points are included the picture is different from Figure \ref{fig-lowlatency-throughput} and XRootD-HTTPS in average outperforms GridFTP by $~6Gbps$ or around thirty percent. However it can also be seen that both are equally sensitive to latency. This could be explained in part of both using the same TCP congestion protocol underneath.

\begin{figure}[h]
\centering
\frame{\includegraphics[width=8cm,clip]{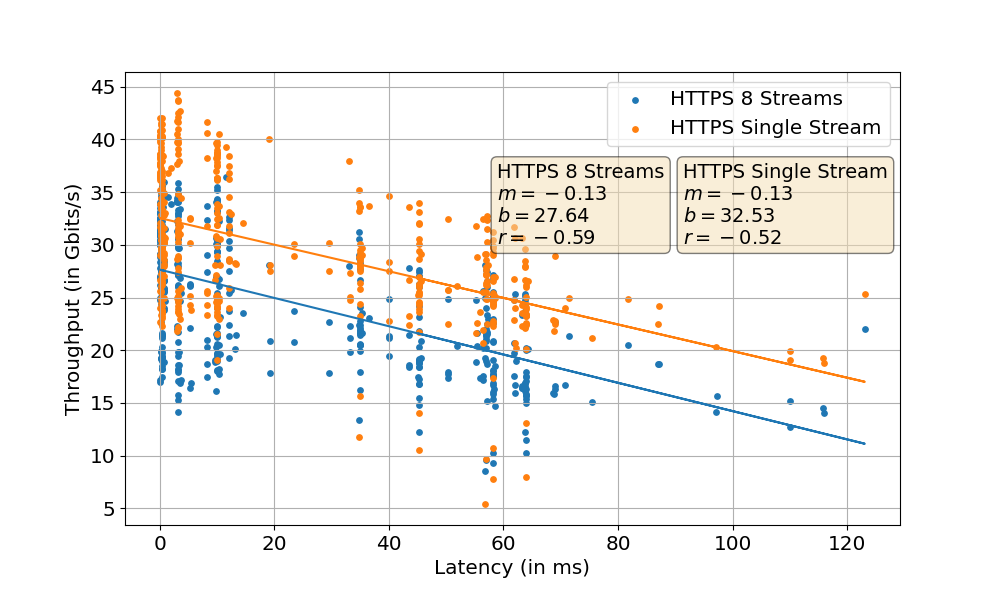}}
\caption{Comparison of throughput in XRootD-HTTPS single stream and multi-stream TPCs}
\label{fig-https-single-stream-comp}
\end{figure}

Finally since the code implementation differs for single stream and multi-stream, it is interesting to understand the performance difference between XRootD-HTTPS multi-stream and single stream. Results summarized in Figure \ref{fig-https-single-stream-comp} show that single stream outperforms multi-stream and both share the same sensitivity to latency. 

\section{Conclusions and future work}
\label{sec-conclusions}
Results obtained in this work are consistent with past benchmarking efforts showing that host to host transfer of XRootD-HTTPS TPC are limited at $45Gbps$. But they also show that for the future HL-LHC it will be an improvement to migrate from GridFTP since in average it underperforms its possible replacement. Moreover the results show there is a throughput advantage by using single stream transfers and multi-stream transfers do not improve latency sensitivity. In addition the machinery developed for this work could be used in the future to benchmark any file transfer protocol that supports Third Party Copy. Finally, although the results in this work only involve single host to single host transfers, in practice grid sites set up an array of nodes to perform the transfers both for efficiency and fault tolerance reasons. Future work on this subject should include the limitations when multiple hosts are involved and perform transfers on the same network link.
\section*{Acknowledgement}
The authors would like to thank the different funding agencies for this work, in particular the National Science Foundation through the following grants: OAC-2030508, OAC-1836650, MPS-1148698 and OAC-1541349; and the University of California, San Diego Physical Sciences Department for the Undergraduate Summer Research Award 2020.
\bibliography{TPC.bib}

\begin{thebibliography}{10}

\bibitem{wlcg}
I.~Bird, \emph{Computing for the Large Hadron Collider} (Annual Reviews, 2011),
  Vol.~61, pp. 99--118,
  \urlstyle{tt}\url{http://www.annualreviews.org/doi/abs/10.1146/annurev-nucl-102010-130059}

\bibitem{gridftp}
W.~Allcock, J.~Bresnahan, R.~Kettimuthu, M.~Link, C.~Dumitrescu, I.~Raicu,
  I.~Foster, \emph{The Globus Striped GridFTP Framework and Server}, in
  \emph{Proceedings of the 2005 ACM/IEEE Conference on Supercomputing} (IEEE
  Computer Society, Washington, DC, USA, 2005), SC '05, pp. 54--, ISBN
  1-59593-061-2, \urlstyle{tt}\url{https://doi.org/10.1109/SC.2005.72}

\bibitem{xrootdTPC}
{Adye, T}, {Bockelman, B}, {Ellis, K}, {Freyermuth, O}, {Furano, F}, {Ganis,
  G}, {Hanushevsky, A}, {Ito, H}, {Johnson, I}, {Keeble, O} et~al., EPJ Web
  Conf. \textbf{245}, 04034 (2020)

\bibitem{dorigo2005xrootd}
A.~Dorigo, P.~Elmer, F.~Furano, A.~Hanushevsky, WSEAS Transactions on Computers
  \textbf{1} (2005)

\bibitem{osg}
R.~Pordes, D.~Petravick, B.~Kramer, D.~Olson, M.~Livny, A.~Roy, P.~Avery,
  K.~Blackburn, T.~Wenaus, F.~Würthwein et~al., Journal of Physics: Conference
  Series \textbf{78}, 012057 (2007)

\bibitem{ethr}
Microsoft, \emph{ethr}, \url{https://github.com/microsoft/ethr} (2019)

\bibitem{chep2019}
{Fajardo, Edgar}, {Bockelman, Brian}, {Wuerthwein, Frank}, EPJ Web Conf.
  \textbf{245}, 04025 (2020)

\bibitem{tpc-bench}
R.~{Aashay Arora}, \emph{aaarora/tpc-benchmarking} (2021),
  \urlstyle{tt}\url{http://doi.org/10.5281/zenodo.4433039}

\bibitem{osg-ca-gen}
B.~Lin, M.~Selmeci, A.~Arora, M.~Coatsworth.,
  \emph{opensciencegrid/osg-ca-generator} (2021),
  \urlstyle{tt}\url{http://doi.org/10.5281/zenodo.4415845}

\bibitem{maddash}
perfSONAR, \emph{Maddash}, \url{https://github.com/perfsonar/maddash/} (2019)

\end{thebibliography}
\end{document}